%% file: article.tex
\documentclass[12pt]{article}

\usepackage{epsfig}
\usepackage{url}
\usepackage{hyperref}

\input econfmacros.tex
\def\Title#1{\begin{center} {\Large {\bf #1} } \end{center}}

\begin{document}

\Title{Combined Higgs boson measurements and their interpretations with the ATLAS experiment}

Presented at DIS2023: XXX International Workshop on Deep-Inelastic Scattering and Related Subjects, Michigan State University, USA, 27-31 March 2023.

\bigskip\bigskip


\begin{raggedright}

{\it Marc Escalier\index{Escalier, M.}, on behalf of the ATLAS Collaboration\footnote{Copyright CERN for the benefit of the ATLAS Collaboration. CC-BY-4.0 license}\\
IJCLab, PHE department\\
15 Rue Georges Clemenceau\\
91400 Orsay, FRANCE}
\end{raggedright}\\

Abstract: Combining measurements of many production and decay channels of the observed Higgs boson allows for the highest possible measurement precision for the properties of the Higgs boson and its interactions. These combined measurements are interpreted in various ways; specific scenarios of physics beyond the Standard Model (SM) are tested, as well as a generic extension in the framework of the Standard Model Effective Field Theory. The latest highlight results of these measurements and their interpretations performed by the ATLAS \cite{ATLAS:2008xda} Collaboration are discussed.



\section{Production modes and decay channels of the Higgs boson, kinematic properties}

The $H\rightarrow\gamma\gamma$, $H\rightarrow ZZ^*$, $H\rightarrow WW^*$, $H\rightarrow Z\gamma$, $H\rightarrow bb$, $\rightarrow\tau\tau$, $H\rightarrow\mu\mu$ analyses are combined~\cite{ATLAS:2022vkf} in order to probe main production modes
and decay channels. With this Run 2 combined publication, the $WH$, $ZH$, $ttH+tH$ are observed respectively with a significance of~$5.8$, $5.0$ and $6.4$. An observed $95\ \%$ Confidence Level (CL) limit of $15\times SM$ is obtained on $tH$. On the decay side, the $H\rightarrow bb$ mode is observed with a significance of $7.0$. The observed significances of $H\rightarrow\mu\mu$ and $H\rightarrow Z\gamma$ are respectively of $2.0$ and $2.3$. Coupling strength modifiers are probed using the $\kappa$ framework~\cite{LHCHiggsCrossSectionWorkingGroup:2012nn}, without $Z\gamma$ nor $\mu\mu$ channels, using three scenarios increasingly reducing the assumptions: (1) boson/fermion symmetry, probing $Hff$ and $HVV$, (2) independent $\kappa$ parameters, allowing to observe the compatibility of scaling of coupling with mass of particles, probing also the specific $\kappa_c$ coupling strength modifier by adding the $H\rightarrow cc$ channel, with a limit of $5.7\times SM$, and (3) a generic parametrization with or without invisible and undetected particles, which adds the $H\rightarrow invisible$ channel as input. In this last scenario that allows non-SM particles, couplings of loops are taken as effective instead of resolving them as functions of their fundamental constituents. Non-SM Higgs branching ratios to invisible particles and branching ratios to undetected particles for which the inputs have no sensitivity are disentangled by a constraint of Higgs coupling to vector boson below or equal~1. Kinematic properties of Higgs production are probed using the STXS framework stage 1.2 \cite{Monti:2803606} that splits phase space in so-called truth bins of production modes, kinematics of the Higgs, kinematics of the jets and vector boson and multiplicity of jets. All measurements are compatible with the predictions from the SM.\\

Using a slightly reduced number of channels in the combination (mostly without the $H\rightarrow cc$ channel), the coupling strength modifiers are translated into contour limits in the plane $\tan{\beta}$ vs $\cos{(\beta-\alpha)}$ of the four 2HDM models that prevent tree-level Flavor-Changing Neutral Currents (FCNC). Results are compatible \cite{ATLAS:2021vrm} with the alignment limit $\cos{(\beta-\alpha)}=0$ where the coupling of lightest Higgs boson matches those of the SM one. Non excluded petal regions remain for three types, corresponding to Yukawa coupling $Hff$ of the same magnitude as in the SM but with a different sign.

\section{Interpretation of production and decay of the Higgs boson in SMEFT}

The measurement from ATLAS on Higgs production ($H\rightarrow\gamma\gamma$, $H\rightarrow ZZ^*$, $H\rightarrow WW^*$, $\rightarrow\tau\tau$, $H\rightarrow bb$) and electroweak differential cross-section, as well as results from LEP and SLD experiments on electroweak measurements are combined \cite{ATLAS:2022xyx} to probe either direct Wilson coefficients  ($c_i$) or eigenvectors of them.
The yields are parametrized as functions of $c_i$. It is related to the perturbative development of the Lagrangian where Wilson coefficients weight new operators of energy dimension~6. The SM operators correspond to dimension 4. Dimensions 5 and 7 are ignored from arguments of avoiding violation of baryonic and leptonic numbers. Linear terms (in $c_i$) correspond to interference between the SM and dimension 6 terms, quadratic terms correspond to pure dimension 6 terms. Dimension 8 are neglected but comparison between results from linear and linear+quadratic provides a qualitative estimation of the validity of neglecting them. In order to reduce the high number of operators probed, a symmetry is introduced, imposing same operators within the two first generations of quarks and within the three generations of charged leptons.\\

The relative impact of each Wilson coefficient on the cross-section could be visualized for various STXS truth bins and compared to the corresponding relative uncertainty of each cross-section measurement. In the yield parametrization, the signal acceptance times efficiency, signal shape and background modelisation are assumed to be independent of EFT parameters. This is a good approximation coming from the similar kinematics behaviour in a given truth bin. This is no more true for Higgs decays. For example, in the $H\rightarrow 4\ell$ channel, the subleading dilepton pair distribution varies a lot from the SM when Wilson coefficients are turned on. Branching ratios are thus parametrized as functions of $c_i$. In the case of $H\rightarrow 4\ell$, it is parametrized after applying the fiducial cuts of the analysis to take into account the important change.\\

The parametrization of yields is obtained from a basis of samples with various values of $c_i$. Tree level/LO processes are simulated with Madgraph and SMEFTSim while loop diagrams in production and decay are treated with SMEFT@NLO. The specific $H\rightarrow\gamma\gamma$ is parametrized using an analytic computation. Higher order corrections are obtained and factorized from the SM. Common systematics are treated correlated among the channels.\\

The results are presented in the form of measurements on individual and eigenvector of Wilson coefficients. These last ones are obtained from a Principal Component Analysis based on the Hessian matrix. Results are provided for ATLAS only inputs as well as for the further combination with LEP and SLD. No deviation with respect to the SM is observed.

\section{Fiducial and differential cross-section}

The fiducial and differential cross-sections combination \cite{ATLAS:2022qef} uses $H\rightarrow\gamma\gamma$, $H\rightarrow ZZ^*\rightarrow 4\ell$. The measured fiducial cross-sections from this $\sqrt{s}=13$ TeV data-taking and from the former $\sqrt{s}=7$ TeV and $\sqrt{s}=8$ TeV campaigns are in agreement with the energy depencence prediction.
Five differential cross-sections are measured: multiplicity of jets and $p_T$ of the leading jet, which are sensitive to theory modelization of the QCD radiation and production modes, rapidity of the Higgs $|y_H|$, sensitive to parton density functions of the proton, transverse momentum $p_T^H$ of the Higgs boson, sensitive to perturbative QCD.
The measurements are dominated by statistical uncertainties. The experimental and theoretical uncertainties are correlated among channels. The measurements are compatible among channels.\\

$Hcc$ and $Hbb$ Yukawa couplings are sensitive to this last variable by subleading effects in the loop diagrams, such as gluon fusion and decay of Higgs to pair of photons, thus affecting more generally branching ratio of $H\rightarrow ZZ^*\rightarrow 4\ell$, but also in quark initiated production. In the scenario where both the normalization and shape are exploited, this gives the observed confidence interval on $\kappa_b$ and $\kappa_c$ respectively of $[-1.09, -0.86]\ U\ [0.81, 1.09]$ and $[-2.27, 2.27]$. The channels are combined further with the channels involving directly these couplings: $H\rightarrow bb$, $H\rightarrow cc$ with $VH$ production with or without allowing decay to Beyond-Standard-Model (BSM) particles. This gives most stringent constraints on $\kappa_c$.

\section{Higgs to invisible}

In the SM, decay of Higgs to invisible appears at the branching ratio level of $0.1\ \%$ from the $H\rightarrow 4\nu$ channel. 
More generally, in BSM scenario, Higgs could decay to invisible particle and act as a portal between a dark sector and the SM. 
The signature of the search is the presence of missing transverse energy and a tagging particle. This drives to the combination of $H+j$, $VBF$, $VBF+\gamma$ 
(where selection of these two last ones are orthogonal), $ZH$, $ttH$. For Run 2, most systematics are correlated among channels. 
Since the performance of object reconstruction and identification are made from data-driven measurements, while detector layout and data-taking conditions varied between Run 1 and 2, 
the systematics on objects are uncorrelated. The systematics on background modelling are also uncorrelated to take into account the improvement in the MC simulation and theory prediction.\\

The $95\ \%$ CL observed (expected) limit on the branching ratio of Higgs to invisible is \cite{ATLAS:2023tkt} $0.107$ ($0.077$), driven mostly by $VBF$ and $ZH$ channels. From an effective field theory framework, limits are translated into an spin-independent scattering cross-section of a Weakly Interacting Massive Particle (WIMP) to nucleon, giving complementary indirect limits to those from direct measurement experiments, for scenarios with scalar, majorana, vector of the WIMP.


\section{Probing Higgs self-coupling: $HH+H$}

The Higgs self-coupling $\lambda$ from the Higgs potential, and more generally its measured ratio $\kappa_{\lambda}=\lambda/\lambda_{SM}$ with respect to the SM
could be measured directly from double Higgs production but also indirectly from higher order electroweak corrections to the single Higgs production. In context of such a double Higgs production, the secondary production mode of $VBF\ HH$ introduces a coupling $HHVV$, related to the coupling strength modifier $\kappa_{2V}$, whose double Higgs production is uniquely sensitive to.
The inputs to the combination \cite{ATLAS:2022jtk} are made of three double Higgs channels: $bb\gamma\gamma$, $bb\tau\tau$, $bbbb$ and five single Higgs channels: $\gamma\gamma$, $ZZ^*\rightarrow 4\ell$, $\tau\tau$, $WW^*\rightarrow e\nu\mu\nu$, $bb$. The potential overlap among channels are checked to be negligible, apart for the $ttH$ category of single Higgs populating the $HH\rightarrow bb\tau\tau$ channel. For this reason, this specific category is removed from the $HH+H$ combination.
In the statistical model, the systematics from common physics objects and mass of the Higgs are correlated among channels. Ingredients from different methodologies are uncorrelated. Theoretical systematics are correlated, but those on the single Higgs and double Higgs production are uncorrelated.\\

The combination of $HH$ channels establishes an observed (expected) $95\ \%$ CL limit on the signal of $2.4\ (2.9) \times SM$. The limits on the rate of inclusive signal and on $VBF\ HH$ are translated in a limit on the couplings strength modifiers $\kappa_{\lambda}$ and $\kappa_{2V}$. The combination with the single Higgs allows to relax assumption on Higgs couplings. In a scenario with $\kappa_t$ floating, the confidence intervals are the same as without floating it. A generic scenario floating $\kappa_t$, $\kappa_V$, $\kappa_b$, $\kappa_{\tau}$ is also probed.

\section{Conclusion}

Results presented here improve further those from previous edition of DIS conference. No deviation with respect to the SM is observed.




\bibliographystyle{hep}
\bibliography{article}












 
\end{document}

%% file: econfmacros.tex
\def\beq{\begin{equation}}
\def\eeq#1{\label{#1}\end{equation}}
\def\eeqn{\end{equation}}

\def\beqa{\begin{eqnarray}}
\def\eeqa#1{\label{#1}\end{eqnarray}}
\def\eeqan{\end{eqnarray}}

\let\bar=\overbar

\def\Dslash{\not{\hbox{\kern-4pt $D$}}}
\def\dslash{\not{\hbox{\kern-2pt $\del$}}}

\def\msb{{\bar{\ssstyle M \kern -1pt S}}}

